\documentclass[12pt]{JHEP3}

\usepackage[pdftex]{graphicx}
\usepackage{epsfig,multicol,amsmath}
\newcommand\fverb{\setbox\pippobox=\hbox\bgroup\verb}
\newcommand\fverbdo{\egroup\medskip\noindent%
\fbox{\unhbox\pippobox}\ }			
\newcommand\fverbit{\egroup\item[\fbox{\unhbox\pippobox}]}
\newbox\pippobox
\def\d2bar{$\overline{\mbox D2}$}
\title{Oblique DLCQ M-theory and Multiple M2-branes}
\author{Jin-Ho Cho$^{1}$ and Sunyoung Shin$^{2}$\\
$^{1}$Department of Physics \& Research Institute for Natural Sciences\\
Hanyang University, Haengdang-dong 1, Seongdong-gu, Seoul 133-791, Korea\\
$^{2}$Department of Physics, Sungkyunkwan University\\
Chunchun-dong 300, Jangan-gu, Suwon 440-746, Korea\\\\
E-mail: \email{cho.jinho@gmail.com}, \email{sihnsy@skku.edu}}

\abstract{We propose an oblique DLCQ as a limit to realize a theory of multiple M$2$-branes in M(atrix)-theory context. The limit is a combination of an infinite boosting of a space-like circle and a tuned tilting of the circle direction. We obtain  a series of supergravity solutions describing various dual configurations including multiple M$2$-branes. For an infinite boosting along a circle wrapped obliquely around a rectangular torus, Seiberg's DLCQ limit distorts the torus modulus. In the context of supergravity, we show explicitly how this torus modulus of $\widetilde{\text M}$-theory is realized as the vacuum modulus of dual IIB-theory.}  


\begin{document} 

\section{Introduction}
\subsection{M2-brane Descriptions So Far}

Despite much effort to formulate M-theory, it is yet far from our understanding. Regularization of  supermembrane theory on the light front results in a proposal of M-theory as the supersymmetric matrix quantum mechanics \cite{de Wit:1988ig}. However, as a first quantized theory, it suffers from instability of generating spikes on membranes\cite{de Wit:1988ct}, thus, we regard it just as a second quantized theory of partons (D$0$-branes) \cite{Townsend:1995af}\cite{Banks:1996vh}. Covariant regularization was not satisfactory because the $\kappa$-symmetry fixing leads only to a $10$-dimensional Lorentz symmetric theory \cite{Fujikawa:1997jt}\cite{Fujikawa:1997dx}. 

The recently proposed theory of multiple M$2$-branes has attracted huge interests in this regard \cite{Gustavsson:2007vu}\cite{Bagger:2007jr}\cite{Bagger:2007vi}. This world-volume theory of multiple M$2$-branes passed the basic requirements of being $\mathcal{N}=8$ superconformal theory with SO$(8)$ R-symmetry. However, we do not yet have a manageable representation of $3$-algebra, an inevitable element of the theory.  

\subsection{the Issue in this Paper} 

In this paper, we return to the matrix description of M-theory and see where the multiple M$2$-branes enter especially in the discrete light cone quantization (DLCQ) prescription (leading to a finite $N$ matrix model) \cite{Susskind:1997cw}. Though M$2$-branes, being composed of $N$ D$0$ `partons', have been discussed as solutions\footnote{The solutions break the supersymmetries by half.} of matrix quantum mechanics \cite{Aharony:1996bh}\cite{Lifschytz:1996rw}\cite{Polchinski:1997pz}, what we want to check in this paper is {\it whether the matrix M-theory is dual to a theory of multiple M$2$-branes in some limit} as it is to D$2$-branes. 

DLCQ M-theory on a torus $T^{p}$ is dual to other theories describing various branes \cite{Seiberg:1997ad}\cite{Sen:1997we}. For example, compactified on a transverse $T^{2}$, DLCQ M-theory becomes dual to the $(2+1)$-dimensional super Yang-Mills (SYM) theory describing multiple D$2$-branes. This feature persists until $p=3$, over which the dual theories become strongly coupled and pertain to $11$-dimensions. 

In the conventional DLCQ prescription, the string coupling $g_{s}$ is proportional to $l^{3-p}_{s}$ \cite{Seiberg:1997ad}. Therefore, the theory of D$2$-branes, that is $p=2$ case with its coupling not large enough, cannot be promoted to that of M$2$-branes. 

\subsection{Our Strategy}

To realize multiple M$2$-branes in DLCQ prescription, one has to go to the strong coupling regime of multiple D$2$-branes. One possible way for this is to exploit the M/IIB duality \cite{Schwarz:1995dk}\cite{Aspinwall:1995fw}\cite{Schwarz:1995jq}. (See Refs.~\cite{Dai:2003dy}\cite{Dai:2004ke}  for a detailed analysis on the issue performed in the matrix theory context.) More specifically, this duality between M-theory on a torus $T^{2}$ and IIB-theory on a circle $S^{1}$ asserts that the torus moduli of M-theory is the same as the vacuum moduli of IIB-theory. By a complex structure deformation on the torus of M-theory, one can reach multiple $(p, q)$-strings in IIB-theory, which could result in the strong coupling in IIA-theory via T-duality.

The tool we employ to achieve our goal is a variant of DLCQ. The oblique DLCQ is to tilt the momentum direction of an M-wave, off the M-circle, i.e., the direction to be compactified. We will show that Seiberg's rescaling used to reach $\widetilde{\text M}$-theory \cite{Seiberg:1997ad} provides the desired complex structure deformation of the torus. The directions transverse to the tilted momentum direction will shrink to deform the torus shape. (The over-tilde stands for a different characteristic length $\tilde{l}_{p}$ from that of the original M-theory.)

\subsection{the Organization of this Paper}

We organize this paper as follows. In the next section, we will recapitulate briefly the basic idea of DLCQ prescription. We will see how the idea of relating the large but nearly lightlike circle with the small spacelike circle, comes in our setup, M-theory on $T^{3}$ (including the M-circle direction). Subsequently in Sec. \ref{seciii}, we will explain how the oblique DLCQ procedure deforms the complex structure of the torus $T^{2}$, a section of $T^{3}$. This leads to an $\widetilde{\text M}$-theory on a slanted $3$-torus. 

Our basic strategy is to go to the dual description well-suited for the small spacelike circle limit. In Sec. \ref{seciv}, we first go to a $\widetilde{\text{IIA}}$-theory ($\widetilde{\text M}$-theory on a small spacelike circle) compactified on a $2$-torus. The finite $N$ momentum sector of the original M-theory corresponds to $N$ units of (D$0+$momentum) bound state. Since the torus size is very small, we go over to another IIA-theory on a large torus via a IIB-theory by sequential T-duality transformations. Sec. \ref{secv} concerns the IIB configuration. It turns out to be $N$ units of $(p,\, q)$-strings. We explicitly show that the vacuum modulus of this IIB-theory coincides with the torus modulus of $\widetilde{\text M}$-theory. This result confirms the duality between both theories in the context of supergravity solutions. Originally, it was shown by comparing the BPS spectra of both theories \cite{Schwarz:1995dk}. Being back to IIA-theory in Sec. \ref{secvi}, we have a non-threshold bound state of D$2$-F$1$-branes. In Sec. \ref{secvii}, we estimate the order of the string coupling in the size $\tilde{R}_{s}$ of small spacelike circle. It diverges as $g_{IIA}\sim \mathcal{O}(\tilde{R}^{-1/4}_{s})$. This justifies M-lifting of the configuration. In Sec. \ref{secviii}, we eventually reach a multiple M$2$-brane configuration. It implies that the oblique DLCQ M-theory on $T^{2}$ is dual to the theory of multiple M$2$-branes\footnote{When we say DLCQ M-theory on $T^{p}$, we mean the M-theory on $T^{p+1}$ with one of the circle directions to be lightlike.}. We also show how the extended U-duality transformation is realized in the supergravity context. Sec.~\ref{secix} concludse this paper with some remarks. We compare the oblique DLCQ scheme with the conventional DLCQ procedure. We also specify the parameters involved in the oblique DLCQ scheme and discuss the decoupling limit. For convenience, the appendix collects various sizes of the compact directions introduced throughout this paper. 

\section{DLCQ in Brief}

\subsection{M-theory on a Finite Lightlike Circle}

For our notation setup, this section recapitulates the prescription of DLCQ as was presented in Ref. \cite{Seiberg:1997ad}. This idea of DLCQ will be exploited frequently in the forthcoming parts of this paper.   

DLCQ M-theory \cite{Susskind:1997cw} follows the spirit of BFSS conjecture (advocated by  Bank, Fischler, Shenker, and Susskind \cite{Banks:1996vh}). BFSS proposed that the uncompactified M-theory in the infinite momentum frame (IMF) is equivalent to the large $N$ limit of a matrix quantum mechanics of D$0$-branes. This non-perturbative definition of uncompactified M-theory in IMF can be generalized to the finite $N$ matrix quantum mechanics, but this time, it is equivalent to the M-theory compactified on a lightlike circle of a finite radius \cite{Susskind:1997cw}. 

The question concerning DLCQ M-theory is why the minimal super Yang-Mills matrix quantum mechanics is enough to represent the strongly coupled theory. Conventionally we have to include higher derivative terms in the strong coupling regime.  

Refining the lightlike circle idea of DLCQ M-theory, Seiberg gave us the answer to this question \cite{Seiberg:1997ad}. If we replace the lightlike circle with a nearly lightlike circle, clearly the M-theory on a large nearly lightlike circle is related with the M-theory on a small spacelike circle via a boosting. Let us consider a spacelike circle specified by the identification relation $(T,\,X_{11})\sim (T,\,X_{11}-2\pi \tilde{R}_{s})$.  The parameter $\tilde{R}_{s}$ denotes the radius of the spacelike circle. By the boosting, 
\begin{eqnarray}\label{boost}
\left(\begin{array}{c}T \\X_{11}\end{array}\right) &=&\left(\begin{array}{ccc}\cosh{\gamma}\, & \sinh{\gamma}\,  \\\sinh{\gamma}\, & \cosh{\gamma}\, \end{array}\right)\left(\begin{array}{c}T' \\X'_{11}\end{array}\right), 
\end{eqnarray}
with the boosting parameter $\gamma$ given by
\begin{equation}\label{boostparameter}
\tanh{\gamma}\,=\frac{R}{\sqrt{R^{2}+2\tilde{R}^{2}_{s}}},
\end{equation}
one can get a nearly lightlike circle; 
\begin{equation}
(T',\,X'_{11})\sim (T'+\frac{2\pi R}{\sqrt{2}},\,X'_{11}-2\pi\sqrt{\frac{R^{2}}{2}+\tilde{R}^{2}_{s}}).
\end{equation}
More specifically in the lightcone coordinates $X'^{\pm}=(T'\pm X'_{11})/\sqrt{2}$, the identification is recast as 
\begin{equation}\label{nulldirection}
(X'^{+},\,X'^{-})\sim (X'^{+}-\frac{\pi \tilde{R}^{2}_{s}}{R},\,X'^{-}+2\pi R+\frac{\pi \tilde{R}^{2}_{s}}{R}).
\end{equation}
In $\tilde{R}_{s}/R\rightarrow 0$ limit, it describes a nearly lightlike compactification of radius $R$ along $X'^{-}$.

\subsection{Seiberg's Limit}

Another ingredient necessary for understanding DLCQ M-theory is that we have to rescale the Planck length. Otherwise, the M-theory on a small spacelike circle will be reduced to a ten-dimensional theory of weakly coupled but tensionless strings. Indeed, the ten-dimensional string length and the string coupling are given \cite{Witten:1995ex} by
\begin{equation}
\frac{1}{l^{2}_{s}}=\frac{\tilde{R}_{s}}{l^{3}_{p}},\qquad g^{2}_{s}=\left(\frac{\tilde{R}_{s}}{l_{p}} \right)^{3}.
\end{equation}    
For fixed $l_{p}$, both the string tension and the string coupling vanish as $\tilde{R}_{s} \rightarrow 0$. What is worse is that the lightcone energy $P'^{-}\sim R /l^{2}_{p}$, is related to the value $P^{-}\sim \tilde{R}_{s}/l^{2}_{p}$ by the boosting \eqref{boost}, thus $P^{-}$ vanishes in the same limit. 

To avoid this bizarre behavior, Seiberg suggested that we introduce a new scale $\tilde{l}_{p}$ so that we focus on the mode of a fixed value of $P^{-}$. We keep the value,
\begin{equation}\label{energyfix}
\frac{\tilde{R}_{s}}{\tilde{l}^{2}_{p}}=\frac{R}{l^{2}_{p}},
\end{equation}
finite in the limit of $\tilde{R}_{s}\rightarrow 0$.
This newly introduced scale affects the other compact directions too, but one can control those other compact directions by redefining the numberings on the corresponding axes. It implies that
\begin{equation}\label{numberingfix}
\frac{\tilde{R}_{i}}{\tilde{l}_{p}}=\frac{R_{i}}{l_{p}}
\end{equation}
will be kept finite. Here, $i=1,\,2,\cdots p$ for $p$ compact directions other than the M-circle direction $X_{11}$.
 
The $\widetilde{\text M}$-theory (the eleven dimensional theory with the new Planck length $\tilde{l}_{p}$) on the small spacelike circle becomes $\widetilde{\text{IIA}}$-theory whose coupling and the string length are small, thus well-defined. Indeed,
\begin{eqnarray}
\tilde{g}_{s}&=&\left(\frac{\tilde{R}_{s}}{\tilde{l}_{p}} \right)^{\frac{3}{2}}=\tilde{R}^{\frac{3}{4}}_{s}\left(\frac{R}{l^{2}_{p}} \right)^{\frac{3}{4}},  \nonumber\\
\tilde{l}^{2}_{s}&=&\frac{\tilde{l}^{3}_{p}}{\tilde{R}_{s}}=\tilde{R}^{\frac{1}{2}}_{s}\left( \frac{l^{2}_{p}}{R}\right)^{\frac{3}{2}}. 
\end{eqnarray}
 
 For compact directions, we have to take T-dualities because the sizes of those directions,
\begin{equation}
\tilde{R}_{i}=R_{i}\sqrt{\frac{\tilde{R}_{s}}{R}},
\end{equation} 
become very small in $\tilde{R}_{s}\rightarrow 0$ limit. Under T-dualities, the string coupling transforms as
\begin{equation}
\tilde{g}'_{s}=\tilde{R}^{\frac{3-p}{4}}_{s}\left(\frac{R}{l^{2}_{p}} \right)^{\frac{3(p+1)}{4}}\prod^{p}_{i=1}\Sigma_{i},\qquad \Sigma_{i}=\frac{\tilde{l}^{2}_{s}}{\tilde{R}_{i}}=\left(\frac{l_{p}}{R_{i}} \right)\left(\frac{l^{2}_{p}}{R} \right).
\end{equation}
Here, the radii $\Sigma_{i}$ denote the dual circle sizes. Unless $p>3$, the coupling is finite and the corresponding world-volume theory describes $N$ stacks of D$p$-branes.

\section{an Oblique Torus}\label{seciii}
In the previous section, we observed that DLCQ M-theory on $T^{p}$ is dual to a theory describing multiple D$p$-branes, when $p\leq 3$. Especially for $p=2$, the weak string coupling $\tilde{g}'_{s}\sim \mathcal{O}(\tilde{R}^{1/4}_{s})$ implies that the dual theory describes just D$2$-branes but not multiple M$2$-branes.

 In order to locate multiple M$2$-branes in DLCQ M-theory, we modify the setup for the case of $T^{2}$ so that it is dual to a theory of multiple D$2$-branes but with its coupling strong. Therefore we have to consider some modification of DLCQ procedure that results in the S-duality effect at some point on the chain of T-dualities. This might then promote the string coupling to be very strong at the final stage of IIA-theory involving D$2$-branes, so that we can go up to eleven dimensions. 
 
 The idea is to exploit the duality between M-theory on $T^{3}$ (including M-circle) and IIB-theory on $T^{2}$ \cite{Schwarz:1995jq}. We will consider DLCQ M-theory on a slanted torus. Since the torus modulus  of M-theory corresponds to the vacuum modulus of IIB-theory, the corresponding IIB-theory will have a non-trivial vacuum attainable from a trivial one by an S-duality.  

\subsection{a Wave on a Tilt}

Suppose a rectangular $3$-torus whose coordinates are identified as 
\begin{equation}\label{eq31}
x_{11}\sim x_{11}+2\pi r_{11},\quad x_{1}\sim x_{1}+2\pi r_{1},\quad x_{2}\sim x_{2}+2\pi r_{2}.
\end{equation}
We consider a uniform wave over $(x_{11},\,x_{1})$-plane, propagating along a generic direction tilted by an angle $\theta$ with respect to the axis $x_{11}$. The wave, being compatible with the torus periods, takes the form
\begin{equation}\label{wave1}
\psi_{\vec{k}}(x_{11},\,x_{1})\sim \exp{\frac{i}{r_{11}}\left(m x_{11}+\frac{1}{\tau_{2}}n x_{1} \right)}.
\end{equation}
Here, the modulus of the rectangular torus is $\vec{\tau}=(0,\,r_{1}/r_{11})$. 

Unless $\theta=0$ or $\pi/2$, this gives the relation between the mode numbers $m$ and $n$, given the radii $r_{11}$ and $r_{1}$.
Since the wave propagates with the momentum, $\vec{k}=(m/r_{11},\,n/(r_{11}\tau_{2}))$,  the angle $\theta$ can be specified by the relation; 
\begin{equation}\label{direction}
\tan{\theta}\,=\frac{n }{m \tau_{2}}=\frac{nr_{11}}{mr_{1}}.
\end{equation}
If $\theta=0$, the wave propagates along $x_{11}$-direction and $n=0$, which corresponds to the conventional DLCQ discussed in the literatures \cite{Seiberg:1997ad}\cite{Sen:1997we}. When $\theta=\pi/2$, the propagating direction is along $x_{1}$-direction and $m=0$. In this paper, we assume $0<\theta< \pi/2$, thus $m\ne 0$ and $n\ne 0$. It is also assumed that the mode number $m$ along $x_{11}$-direction is larger than the number $n$ along $x_{1}$-direction. This latter condition is necessary to avoid some singular point in the charge density. We will discuss it in Sec. \ref{secviii}.

Instead of using the radii $r_{11}$ and $r_{1}$, we prefer to use some effective radii $R_{s}$ and $\tilde{R}_{s}$. Upon the compactification of $(x_{11},\,x_{1})$-plane, the Kaluza-Klein wave results in the mass $M$ in lower dimensions;
\begin{equation}\label{eq34}
M^{2}=\frac{1}{r^{2}_{11}\tau^{2}_{2}}\left(m^{2}\tau^{2}_{2}+n^{2} \right)=\frac{m^{2}}{r^{2}_{11}\cos^{2}{\theta}\,}=\frac{n^{2}}{r^{2}_{1}\sin^{2}{\theta}\,}.
\end{equation}
Hence one can regard the wave either as $m$-th Kaluza-Klein mode around a circle of an effective radius $R_{s}\equiv r_{11}\cos{\theta}\,$, or equivalently as $n$-th mode around a circle of another effective radius $\tilde{R}_{s}\equiv r_{1}\sin{\theta}\,$. Written in terms of these radii, the relation \eqref{direction} concerning the propagation direction becomes $n R_{s}=m\tilde{R}_{s}$. 

\subsection{Seiberg's Limit in the Oblique DLCQ}

In order to describe the metric configuration for the wave, it is convenient to use the adapted coordinate $(T,\,X_{11},\,X_{1})$;
\begin{eqnarray}
\left(\begin{array}{c}t \\x_{11} \\x_{1}\end{array}\right)&=&\left(\begin{array}{ccc}1 & 0 & 0 \\0 & \cos{\theta}\, &\!\! -\sin{\theta}\, \\0 & \sin{\theta}\, &\,\,\, \cos{\theta}\,\end{array}\right)\left(\begin{array}{c}T \\X_{11} \\X_{1}\end{array}\right)\,. 
\end{eqnarray}
The wave generates the following configuration;
\begin{equation}\label{metricm}
ds^{2}=-dT^{2}+dX^{2}_{11}+dX^{2}_{1}+dx^{2}_{2}+\left(f(r)-1 \right)\left(dT-dX_{11} \right)^{2}+\sum^{9}_{i=3}dx^{2}_{i}.
\end{equation}
Here, the function $f(r)$ is a harmonic function in the transverse seven spatial directions;
\begin{equation}
f(r)=1+\frac{Q}{r^{5}},\qquad r^{2}=\sum^{9}_{i=3}x^{2}_{i}\,.
\end{equation}

The charge parameter $Q$ concerns the discrete momentum mode number $n$ in the unit of $1/\tilde{R}_{s}$ (or equivalently mode number $m$ in the unit of $1/R_{s}$) along the propagating direction $X_{11}$. By matching the ADM momentum charge \cite{Myers:1986un} with that of the wave \eqref{wave1}, one can note its form
\begin{equation}
Q\sim \frac{l^{9}_{p}}{r_{1}r_{2}r_{11}}\frac{n}{\tilde{R}_{s}}\,=\frac{l^{9}_{p}}{r_{1}r_{2}r_{11}}\frac{m}{R_{s}}\,.
\end{equation}
Therefore, the geometry looks like that of a momentum wave along a compact spatial circle of radius $\tilde{R}_{s}$.

The background geometry of DLCQ M-theory is the Aichelberg-Sexl type metric \cite{Aichelburg:1970dh} that describes a momentum wave along a nearly lightlike circle of radius $R$. (This identification was first discussed in Ref. \cite{Hyun:1997zt}.) By the same boosting as \eqref{boost} (with its boosting parameter \eqref{boostparameter}) the geometry becomes;
\begin{eqnarray}
ds^{2}&=&-2dX'^{+}dX'^{-}+2 \left(f-1 \right)e^{-2\gamma}dX'^{-2} +dX^{2}_{1}+dx^{2}_{2}+\sum^{9}_{i=3}dx^{2}_{i} \nonumber\\
&=&-2dX'^{+}dX'^{-}+2 \frac{Q e^{-2\gamma}}{r^{5}}dX'^{2-} +dX^{2}_{1}+dx^{2}_{2}+\sum^{9}_{i=3}dx^{2}_{i}.
\end{eqnarray}

Now, let us bring the situation to the $\widetilde{\text M}$-theory by introducing a new Planck scale $\tilde{l}_{p}$. Incorporating the scaling conditions \eqref{energyfix} and \eqref{numberingfix}, one can rescale all the coordinates transverse to the direction of wave propagation as
\begin{eqnarray}\label{coordrescale}
T&\rightarrow&\tilde{T}=T\frac{\tilde{l}_{p}}{l_{p}}= T\left(\frac{\tilde{R}_{s}}{R} \right)^{\frac{1}{2}},\quad X_{1}\rightarrow\tilde{X}_{1}=X_{1}\left(\frac{\tilde{R}_{s}}{R} \right)^{\frac{1}{2}},\nonumber\\
&& x_{i}\rightarrow\tilde{x}_{i}=x_{i}\left(\frac{\tilde{R}_{s}}{R} \right)^{\frac{1}{2}} \quad(i=2,\,3,\cdots, 9),
\end{eqnarray}
while keeping the coordinate $X_{11}$ the same, i.e., $\tilde{X}_{11}=X_{11}$.\footnote{One could equivalently choose $R_{s}$ rather than $\tilde{R}_{s}$ in this rescaling. However, our choice of $\tilde{R}_{s}$ would be more convenient for our later use because we will mostly focus on $\theta \rightarrow \pi/2$ limit.} Hereafter we use $\alpha\equiv \tilde{l}_{p}/l_{p}$ interchangeably just for typographical convenience.

\subsection{the Torus Modulus}

\begin{figure}[htbp]
\begin{center}
\includegraphics[width=7cm]{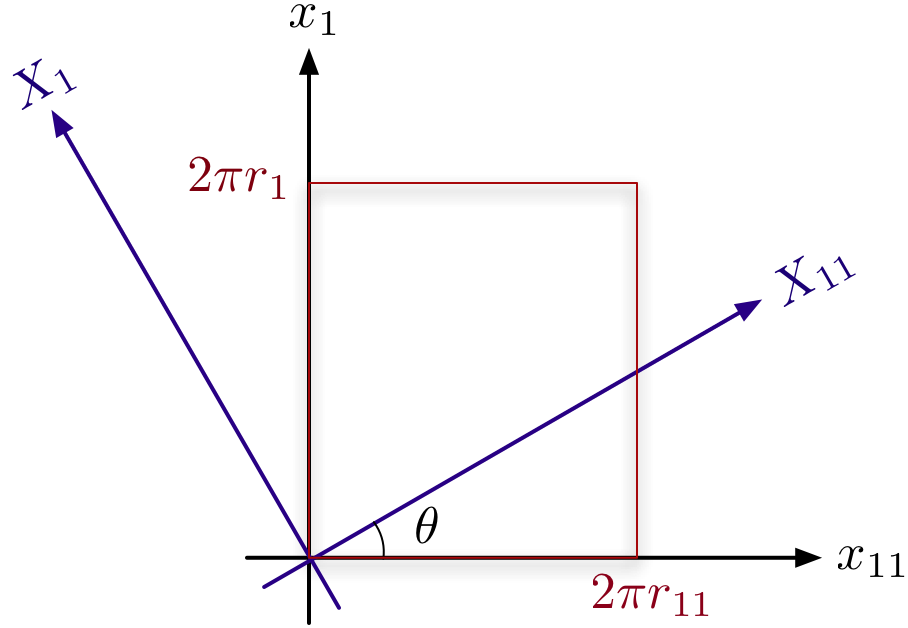}
\includegraphics[width=7cm]{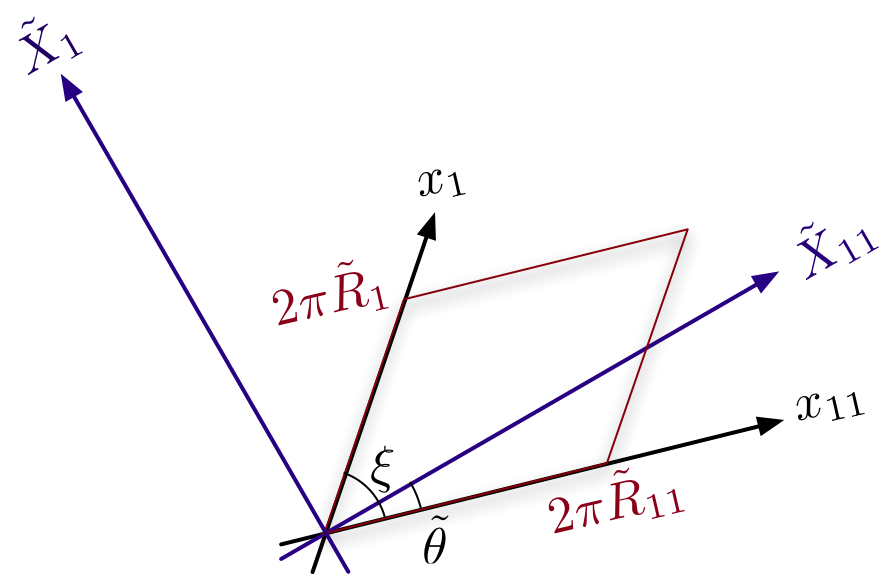}
\caption{The rectangular torus (the left figure) in M-theory is mapped into a slanted torus (the right figure) in $\widetilde{\text M}$-theory. The wave is traveling along $X_{11}$ (or $\tilde{X}_{11}$) with the wave vector $\vec{k}=(m/r_{11},\,n/r_{1})$.}
\label{default}
\end{center}
\end{figure}

The axes $x_{11}$ and $x_{1}$ are deformed so that they are no longer orthogonal in the tilted coordinates. 
The relation between the original coordinates $(x_{11},\,x_{1})$ and the rescaled adapted coordinates $(\tilde{X}_{11},\,\tilde{X}_{1})$ is given by
\begin{eqnarray}
\left(\begin{array}{c}x_{11} \\x_{1}\end{array}\right)&=&\left(\begin{array}{cc}\,\,\,\cos{\theta}\, & -\sin{\theta}\, \\\,\,\,\sin{\theta}\, &\,\,\,\, \,\cos{\theta}\,\end{array}\right)\left(\begin{array}{c}\tilde{X}_{11} \\{\tilde{X}_{1}}/{\alpha}\end{array}\right).  \nonumber\\
\end{eqnarray}
The $x_{11}$-axis ($\sqrt{\tilde{R}_{s}}\tilde{X}_{11} \sin{\theta}\,+\sqrt{R}\tilde{X}_{1}\cos{\theta}\,=0$) and $x_{1}$-axis ($\sqrt{\tilde{R}_{s}}\tilde{X}_{11}\cos{\theta}\,-\sqrt{R}\tilde{X}_{1}\sin{\theta}\,=0$) are intersecting at an angle $\xi$ determined by
\begin{equation}\label{angle}
\tan{\xi}\,=\frac{\alpha}{\cos{\theta}\,\sin{\theta}\,\left( 1-\alpha^{2}\right)}.
\end{equation}
The angle $\theta$ (specifying the propagation direction) is deformed to be $\tilde{\theta}$ satisfying
\begin{equation}\label{tilde}
\tan{\tilde{\theta}}\,=\alpha \tan{\theta}\,.
\end{equation}

The period vectors $(2\pi r_{11},\,0)$ and $(0,\,2\pi r_{1})$ of $(x_{11},\,x_{1})$-frame read in $(\tilde{X}_{11},\,\tilde{X}_{1})$-plane as 
\begin{eqnarray}
(2\pi r_{11},\,0)&\longrightarrow&\left(2\pi r_{11}\cos{\theta}\,,\,-2\pi r_{11}\,\alpha\sin{\theta}\, \right),  \nonumber\\
(0,\,2\pi r_{1})&\longrightarrow&\left(2\pi r_{1}\sin{\theta}\,,\,2\pi r_{1}\,\alpha\cos{\theta}\, \right).
\end{eqnarray}
Therefore, the rescaled periods are
\begin{eqnarray}\label{eq315}
2\pi\tilde{R}_{11}&=&2\pi r_{11}\sqrt{\cos^{2}{\theta}\,+\alpha^{2}\sin^{2}{\theta}\,}=2\pi R_{s}\sqrt{1+\alpha^{2}\tan^{2}{\theta}\,} \nonumber\\
&=&2\pi R_{s}\sec{\tilde{\theta}},\nonumber\\
2\pi\tilde{R}_{1}&=&2\pi r_{1}\sqrt{\sin^{2}{\theta}\,+\alpha^{2}\cos^{2}{\theta}\,}=2\pi \tilde{R}_{s}\sqrt{1+\alpha^{2}\cot^{2}{\theta}}\nonumber\\
&=& 2\pi \tilde{R}_{s}\sqrt{1+\alpha^{4}\cot^{2}{\tilde{\theta}}}.
\end{eqnarray}
All these are illustrated in Fig. \ref{default}. 

Seiberg's rescaling is a moduli transformation that transforms the rectangular torus into a slanted torus.
One can express the torus moduli as
\begin{equation}
\tilde{\tau}=\frac{\tilde{R}_{1}}{\tilde{R}_{11}}e^{i\xi},
\end{equation}
assuming a new orthogonal frame so that the period vector $\vec{T}_{11}$ takes the components $2\pi\tilde{R}_{11}(1,\,0)$. From Eq. (\ref{angle}), we note that
\begin{eqnarray}\label{torusmodulus}
\tilde{\tau}_{1}+i\tilde{\tau}_{2}&=&\frac{r_{1}\left(\cos{\theta}\,\sin{\theta}\, \left(1-\alpha^{2} \right)+i\alpha \right)}{r_{11}\left(\cos^{2}{\theta}\,+\alpha^{2}\sin^{2}{\theta}\,\right)}\nonumber\\
&=&\frac{n \cos^{2}{\tilde{\theta}}\,}{m \tan{\tilde{\theta}}\,}\left(\left(1-\alpha^{2} \right)\tan{\tilde{\theta}}\,+i \left(\alpha^{2}+\tan^{2}{\tilde{\theta}}\, \right) \right). 
\end{eqnarray}

The upshot is that we are considering a wave propagating on a slanted $3$-torus. The directions, $(x_{11},\,x_{1},\,x_{2})$, compose the torus but it is slanted with the modulus \eqref{torusmodulus} in $(x_{11},\,x_{1})$-plane. The wave is of the form
\begin{equation}
\tilde{\psi}_{\vec{\tilde{k}}}(\tilde{X}_{11},\,\tilde{X}_{1})\sim \exp{\frac{im\tilde{X}_{11}}{R_{s}}}=\exp{\frac{in\tilde{X}_{11}}{\tilde{R}_{s}}}.
\end{equation} 

\subsection{Tuning the Propagation}

Since we have one more parameter $\theta$ than the conventional DLCQ description, we have a freedom to tune it. 
We are interested in the limit where $\tan\tilde{{\theta}}\,$ is kept finite while $\alpha \rightarrow 0$. We assume $0\le \tilde{\theta}<\pi/2$ without loss of generality. 

The behavior of the angle $\theta$ in this limit is obviously seen if we write it in terms of $\tilde{\theta}$;
\begin{eqnarray}\label{conversion}
\cos{\theta}\,&=&\frac{\alpha \cos{\tilde{\theta}}\,}{\sqrt{\alpha^{2}\cos^{2}{\tilde{\theta}}\,+\sin^{2}{\tilde{\theta}}\,}},  \nonumber\\
\sin{\theta}\,&=&\frac{\sin{\tilde{\theta}}\,}{\sqrt{\alpha^{2}\cos^{2}{\tilde{\theta}}+\sin^{2}{\tilde{\theta}}\,}}.  
\end{eqnarray} 
As $\alpha \rightarrow 0$, the angle $\theta$ either approaches $0$ (when $\tilde{\theta}=0$), or almost becomes $\pi/2$ (if $0<\tilde{\theta}< \pi/2$).
In the limit, the intersection angle $\xi$ approaches the deformed angle $\tilde{\theta}$ because
\begin{equation}\label{angle2}
\tan{\xi}\,=\frac{\alpha^{2}\cos^{2}{\tilde{\theta}}\,+\sin^{2}{\tilde{\theta}}\,}{\cos{\tilde{\theta}}\,\sin{\tilde{\theta}}\,\left(1-\alpha^{2} \right)} \longrightarrow \tan{\tilde{\theta}}\,.
\end{equation}

\subsection{the Rescaled Geometry}

Under Seiberg's rescaling \eqref{coordrescale}, the metric (\ref{metricm}) becomes, in $\widetilde{\text M}$-theory,
\begin{eqnarray}\label{metrictilde}
d\tilde{s}^{2}&=&-d\tilde{T}^{2}+d\tilde{X}^{2}_{11}+d\tilde{X}^{2}_{1}+d\tilde{x}^{2}_{2}+\left(\tilde{f}(\tilde{r})-1 \right)\left(d\tilde{T}-d\tilde{X}_{11} \right)^{2}+\sum^{9}_{i=3}d\tilde{x}^{2}_{i}.  \nonumber\\
\tilde{f}(\tilde{r})&=&1+\frac{\tilde{Q}}{\tilde{r}^{5}},\qquad (\tilde{r}^{2}=\sum^{9}_{i=3}\tilde{x}^{2}_{i}).  
\end{eqnarray}
The charge parameter now takes the form
\begin{equation}
\tilde{Q}\sim \frac{\tilde{l}^{9}_{p}}{\tilde{r}_{2}\tilde{R}_{1}\tilde{R}_{11}\sin{\xi}\,}\frac{n}{\tilde{R}_{s}}\sim\alpha^{7}Q,
\end{equation}
where the first factor concerns the $8$-dimensional Newton's constant because its denominator, $\tilde{r}_{2}\tilde{R}_{1}\tilde{R}_{11}\sin{\xi}\,$, is the volume of the slanted $3$-torus.

Rewriting the metric (\ref{metrictilde}) in $(t,\,x_{11},\,x_{1})$-coordinates, we get 
\begin{eqnarray}
d\tilde{s}^{2}&=&\left(\tilde{f}\cos^{2}{\theta}\,+\alpha^{2}\sin^{2}{\theta}\, \right)\left(dx_{11}+\frac{\cos{\theta}\,\left(\alpha(1-\tilde{f})dt+(\tilde{f}-\alpha^{2})\sin{\theta}\,dx_{1} \right)}{\tilde{f}\cos^{2}{\theta}\,+\alpha^{2}\sin^{2}{\theta}\,} \right)^{2}  \nonumber\\
&&+\frac{\alpha^{2}\tilde{f}}{\tilde{f}\cos^{2}{\theta}\,+\alpha^{2}\sin^{2}{\theta}\,}\left(dx_{1}+\frac{(1-\tilde{f})\alpha \sin{\theta}\,dt}{\tilde{f}} \right)^{2} -\frac{\alpha^{2}dt^{2}}{\tilde{f}}+\alpha^{2}\sum^{9}_{i=2}dx^{2}_{i},\nonumber
\end{eqnarray}
which is a form ready for the compactification along $x_{11}$-direction.

\section{$\widetilde{\text{IIA}}$: a Bound State of D$0$-branes and Momenta}\label{seciv}

One thing to note regarding the compactification is that, the lower dimensional asymptotic geometry is flat but the coordinates are not Minkowskian. To follow the standard IIA description, we recover the asymptotically Minkowskian coordinates,
\begin{eqnarray}\label{eq41}
&&\tilde{x}_{11}\equiv x_{11}\sqrt{\cos^{2}{\theta}\,+\alpha^{2}\sin^{2}{\theta}\,}=\frac{\alpha x_{11}}{\sqrt{\alpha^{2}\cos^{2}{\tilde{\theta}}\,+\sin^{2}{\tilde{\theta}}\,}},\nonumber\\
&&\tilde{x}_{1}\equiv\frac{\alpha x_{1}}{\sqrt{\cos^{2}{\theta}\,+\alpha^{2}\sin^{2}{\theta}\,}}=x_{1}\sqrt{\alpha^{2}\cos^{2}{\tilde{\theta}}\,+\sin^{2}{\tilde{\theta}}\,},\nonumber\\
&& \tilde{t}=\alpha\,t,\quad \tilde{x}_{i}=\alpha\,x_{i}\quad(i=2,\,3,\cdots,9).
\end{eqnarray}
The coordinate $\tilde{x}_{11}$ is compact as $\tilde{x}_{11}\sim \tilde{x}_{11}+2\pi \tilde{R}_{11}$. The small circle size justifies the compactification to IIA-theory. 

In type IIA language, the configuration looks like
\begin{eqnarray}
ds^{2}_{IIA}
&=&\frac{\tilde{f}}{\sqrt{\tilde{f} \cos^{2}{\tilde{\theta}}\,+\sin^{2}{\tilde{\theta}}\,}}\left(d\tilde{x}_{1}+\frac{1-\tilde{f}}{\tilde{f}}\sin{\tilde{\theta}}\,d\tilde{t} \right)^{2}\nonumber\\
&&+\sqrt{\tilde{f} \cos^{2}{\tilde{\theta}}\,+\sin^{2}{\tilde{\theta}}\,}\left(-\frac{1}{\tilde{f}}d\tilde{t}^{2}+\sum^{9}_{i=2}d\tilde{x}^{2}_{i}\right),\nonumber\\
e^{\phi}&=&
\left(\tilde{f} \cos^{2}{\tilde{\theta}}\,+\sin^{2}{\tilde{\theta}}\, \right)^{\frac{3}{4}}\frac{\tilde{R}_{11}}{\tilde{l}_{s}},\nonumber\\
C^{(1)}
&=& \frac{\cos{\tilde{\theta}}\,}{\tilde{f}\cos^{2}{\tilde{\theta}}\,+\sin^{2}{\tilde{\theta}}\,}\left(\left(1-\tilde{f} \right)d\tilde{t}+\frac{\left(\tilde{f}-\alpha^{2} \right)\sin{\tilde{\theta}}\,}{\alpha^{2}\cos^{2}{\tilde{\theta}}\,+\sin^{2}{\tilde{\theta}}\,}d\tilde{x}_{1} \right),
\end{eqnarray} 
where $(\tilde{R}_{11}/\tilde{l}_{p})^{3}=(\tilde{R}_{11}/\tilde{l}_{s})^{2}$ was used. In $\alpha \rightarrow 0$ limit, the configuration describes $n$ units of D$0+$momentum bound state with the momentum flowing along $\tilde{x}_{1}$-direction. When $\tilde{\theta}=0$, it corresponds to $n$-unit of D$0$-branes, while the configuration becomes that of a momentum wave along $\tilde{x}_{1}$-direction when $\tilde{\theta}$ approaches $\pi/2$. 

\section{$\widetilde{\text{IIB}}: \,(p,\,q)$-strings}\label{secv}

We have still a small compact direction as $\tilde{x}_{1}\sim \tilde{x}_{1}+2\pi \tilde{r}_{1}$. From $r_{1}=\tilde{R}_{s}/\sin{\theta}\,$ and \eqref{conversion}, we note that
\begin{eqnarray}
r_{1}&=&
 \frac{\tilde{R_{s}}\sqrt{\alpha^{2}\cos^{2}{\tilde{\theta}}\,+\sin^{2}{\tilde{\theta}}\,}}{\sin{\tilde{\theta}}\,}. 
\end{eqnarray}
It implies that the size $\tilde{r}_{1}$ shrinks with $\tilde{R}_{s}$;
\begin{eqnarray}\label{eq52}
\tilde{r}_{1}&=&
r_{1}\sqrt{\alpha^{2}\cos^{2}{\tilde{\theta}}\,+\sin^{2}{\tilde{\theta}}\,}=\frac{\tilde{R_{s}}\left(\alpha^{2}\cos^{2}{\tilde{\theta}}\,+\sin^{2}{\tilde{\theta}}\,\right)}{\sin{\tilde{\theta}}\,}.  
\end{eqnarray}

The small circle size $\tilde{r}_{1}$  justifies the T-duality into the IIB configuration.
Taking the T-duality along $\tilde{x}_{1}$-direction, we get
\begin{eqnarray}\label{iibconf}
ds^{2}_{IIB}
&=&\frac{1}{\tilde{f}}\sqrt{ \tilde{f}\cos^{2}{\tilde{\theta}}\,+\sin^{2}{\tilde{\theta}}\,}\left(-d\tilde{t}^{2}+d\bar{x}^{2}_{1} \right)+\sqrt{ \tilde{f}\cos^{2}{\tilde{\theta}}\,+\sin^{2}{\tilde{\theta}}\,}\sum^{9}_{i=2}d\tilde{x}^{2}_{i}, \nonumber\\
e^{\phi_{B}}&=&
\frac{\tilde{f}\cos^{2}{\tilde{\theta}}\,+\sin^{2}{\tilde{\theta}}\,}{\sqrt{ \tilde{f}}}\frac{\tilde{R}_{11}}{\tilde{l}_{s}} \frac{\bar{r}_{1}}{\tilde{l}_{s}}.
\end{eqnarray}
The dual coordinate $\bar{x}_{1}$ is compact as $\bar{x}_{1}\sim\bar{x}_{1}+2\pi\bar{r}_{1}$, where
\begin{eqnarray}\label{dualR1}
 \bar{r}_{1}
 &=&\frac{\tilde{l}^{2}_{s}\sin{\tilde{\theta}}\,}{\tilde{R}_{s}\left(\alpha^{2}\cos^{2}{\tilde{\theta}}\,+\sin^{2}{\tilde{\theta}}\, \right)}.
\end{eqnarray}
The NS-NS field and R-R fields,
\begin{eqnarray}
\bar{B}^{(2)}
&=&-\frac{1-\tilde{f} }{\tilde{f}}\sin{\tilde{\theta}}\,d\tilde{t}\wedge d\bar{x}_{1},\nonumber\\
\bar{C}^{(0)}
&=&\frac{\left(\tilde{f}- \alpha^{2}\right)\tan{\tilde{\theta}}\,}{\left(\alpha^{2}\cos^{2}{\tilde{\theta}}\,+\sin^{2}{\tilde{\theta}}\, \right)\left(\tilde{f}+\tan^{2}{\tilde{\theta}}\, \right)},\nonumber\\
\bar{C}^{(2)}
&=&\frac{\left(\tilde{f}-1 \right)\cos{\tilde{\theta}}\,}{\tilde{f}\cos^{2}{\tilde{\theta}}\,+\sin^{2}{\tilde{\theta}}\,}d\tilde{t}\wedge d\bar{x}_{1},
\end{eqnarray}
describe, in $\alpha \rightarrow0$ limit, $(p,\,q)$-strings along $\bar{x}_{1}$-direction in a non-trivial background of the D-instanton and the dilaton field. 

The vacuum modulus of the IIB configuration coincides with the torus modulus of the $\widetilde{\text M}$-theory.
This was originally proven in Refs.~\cite{Schwarz:1995dk}\cite{Aspinwall:1995fw}\cite{Schwarz:1995jq} by comparing the BPS spectra of both theories. Here, we confirm it in the context of supergravity.
Let us compute the vacuum modulus of the above background. One might naively write it as $\left.(\bar{C}^{(0)}+ie^{-\phi_{B}})\right|_{r \rightarrow\infty}$. However, we should recall that the vacuum modulus in IIB-theory is determined in the canonical frame rather than in the string frame. This correction gives an extra dilatonic factor $e^{-\phi_{B}}|_{r \rightarrow\infty}=g^{-1}_{IIB}$ in every R-R field. Incorporating this factor, we get the following result for the vacuum modulus of the IIB background;
\begin{equation}
\left.(\chi+ie^{-\phi_{B}})\right|_{r \rightarrow\infty}=\frac{r_{1}\left((1-\alpha^{2})\cos{\theta}\,\sin{\theta}\, +i\alpha\right)}{r_{11}(\cos^{2}{\theta}\,+\alpha^{2}\sin^{2}{\theta}\,)}.
\end{equation}
Here, we used $\chi\equiv g^{-1}_{IIB}\bar{C}^{(0)}$ and Eqs. (\ref{iibconf}) and (\ref{dualR1}). This result coincides with the expression (\ref{torusmodulus}) for the torus modulus. Especially one can see that tilting DLCQ direction with respect to the M-circle generates the axion field $\chi$ in type IIB-theory.

\section{Back to $\widetilde{\text{IIA}}$: a Non-threshold D$2$-F$1$ Bound State}\label{secvi}

Since the size of the compact direction in our IIB-configuration is still small as  $\tilde{r}_{2}=\alpha r_{2}$, it is desirable to go back to $\widetilde{\text{IIA}}$-theory via T-duality along $\tilde{x}_{2}$-direction. As its results, we get
\begin{eqnarray}
d\bar{s}^{2}_{IIA'}
&=&\frac{\sqrt{\tilde{f} \cos^{2}{\tilde{\theta}}\,+\sin^{2}{\tilde{\theta}}\,}}{\tilde{f}}\left(-d\tilde{t}^{2}+d\bar{x}^{2}_{1} \right)+\frac{1}{\sqrt{\tilde{f} \cos^{2}{\tilde{\theta}}\,+\sin^{2}{\tilde{\theta}}\,}}d\bar{x}^{2}_{2}\nonumber\\
&&+\sqrt{\tilde{f} \cos^{2}{\tilde{\theta}}\,+\sin^{2}{\tilde{\theta}}\,}\sum^{9}_{i=3}d\tilde{x}^{2}_{i},\\
e^{\bar{\phi}_{A}}
&=&\frac{1}{\sqrt{\tilde{f}}}\left(\tilde{f} \cos^{2}{\tilde{\theta}}\,+\sin^{2}{\tilde{\theta}}\, \right)^{\frac{3}{4}}\frac{\tilde{R}_{11}}{\tilde{l}_{s}}\frac{\bar{r}_{1}}{\tilde{l}_{s}}\frac{\bar{r}_{2}}{\tilde{l}_{s}},
\end{eqnarray}
where the dual radius is given by
\begin{equation}\label{dualR2}
\frac{\bar{r}_{2}}{\tilde{l}_{s}}=\frac{\tilde{l}_{s}}{\tilde{r}_{2}}=\frac{\tilde{l}_{s}}{\alpha r_{2}}.
\end{equation}
(We discern this $\widetilde{\text{IIA}}$-theory on the dual torus $\bar{T}^{2}$ from the initial $\widetilde{\text{IIA}}$ on $T^{2}$ by the subscript `${}_{IIA'}$' or the overbar `$\,\,\bar{}\,\,$' on the variables.)

The NS-NS and R-R fields become
\begin{eqnarray}
\bar{B}^{(2)}
&=&-\frac{1-\tilde{f}}{\tilde{f}}\sin{\tilde{\theta}}\,d\tilde{t}\wedge d\bar{x}_{1}, \nonumber\\
\bar{C}^{(1)}
&=&-\frac{\left(\alpha^{2}-\tilde{f} \right)\tan{\tilde{\theta}}\,}{\left(\alpha^{2}\cos^{2}{\tilde{\theta}}\,+\sin^{2}{\tilde{\theta}}\, \right)\left(\tilde{f}+\tan^{2}{\tilde{\theta}}\, \right)}d\bar{x}_{2},  \nonumber\\
\bar{C}^{(3)}
&=&\frac{\left(\tilde{f}-1 \right)\cos{\tilde{\theta}}\,}{\tilde{f}\cos^{2}{\tilde{\theta}}\,+\sin^{2}{\tilde{\theta}}\,}d\tilde{t}\wedge d\bar{x}_{1}\wedge d\bar{x}_{2}.
\end{eqnarray}
In the limit of $\alpha \rightarrow0$, the configuration describes some D$2$-F$1$ bound state that interpolates D$2$-branes (when $\tilde{\theta}=0$) and F$1$-branes (if $\tilde{\theta}\rightarrow \pi/2$).

\section{Orders of Various Parameters in $\tilde{R}_{s}$}\label{secvii}
The coupling constant $g_{IIA'}=\lim_{r \rightarrow \infty}e^{\bar{\phi}_{A}}$ diverges in the limit of $\alpha \rightarrow 0$. To see this, we use the basic scaling relations used in $\widetilde{\text M}$-theory;
\begin{eqnarray}
\frac{\tilde{R}_{s}}{\tilde{l}^{2}_{p}}&=&\frac{R}{l^{2}_{p}}\equiv A,\quad \frac{\tilde{r}_{2}}{\tilde{l}_{p}}=\frac{r_{2}}{l_{p}}\equiv B_{2}. 
\end{eqnarray}
Here $A$ and $B_{2}$ are some constants of finite quantity.
Therefore
\begin{eqnarray}
\tilde{l}_{p}&=&\sqrt{\tilde{R}_{s}}A^{-\frac{1}{2}},\qquad \tilde{l}_{s}=\tilde{R}_{s}^{\frac{1}{4}}A^{-\frac{3}{4}},  \nonumber\\
\tilde{r}_{2}&=&\alpha r_{2}=\frac{\tilde{l}_{p}}{l_{p}}r_{2}=\sqrt{\tilde{R}_{s}}A^{-\frac{1}{2}}B_{2},  \nonumber\\
\tilde{R}_{11}&=&r_{11}\sqrt{\cos^{2}{\theta}\,+\alpha^{2}\sin^{2}{\theta}\,}
=\frac{R_{s}}{\cos{\tilde{\theta}}\,}=\frac{m\tilde{R}_{s}}{n\cos{\tilde{\theta}}\,},\nonumber\\
\tilde{r}_{1}&=&r_{1}\sqrt{\alpha^{2}\cos^{2}{\tilde{\theta}}\,+\sin^{2}{\tilde{\theta}}\,}=\frac{\tilde{R}_{s}}{\sin{\tilde{\theta}}\,}\left(\alpha^{2}\cos^{2}{\tilde{\theta}}\,+\sin^{2}{\tilde{\theta}}\, \right).
\end{eqnarray}
The factors composing the string couplings take the orders as follow;
\begin{eqnarray}
\frac{\tilde{R}_{11}}{\tilde{l}_{s}}&=&\frac{m \tilde{R}^{\frac{3}{4}}_{s}A^{\frac{3}{4}}}{n\cos{\tilde{\theta}}\,},  \nonumber\\
\frac{\bar{r}_{1}}{\tilde{l}_{s}}&=&\frac{\tilde{l}_{s}}{\tilde{r}_{1}}=\frac{\sin{\tilde{\theta}}\, }{ \tilde{R}^{\frac{3}{4}}_{s}A^{\frac{3}{4}}\left(\alpha^{2}\cos^{2}{\tilde{\theta}}\,+\sin^{2}{\tilde{\theta}}\, \right)},  \nonumber\\
\frac{\bar{r}_{2}}{\tilde{l}_{s}}&=&\frac{\tilde{l}_{s}}{\tilde{r}_{2}}=\frac{1}{\tilde{R}^{\frac{1}{4}}_{s}A^{\frac{1}{4}}B_{2}}.
\end{eqnarray}
This implies that
\begin{eqnarray}
g_{IIB}&=&\frac{m \sin{\tilde{\theta}}\,}{n\cos{\tilde{\theta}}\,\left(\alpha^{2}\cos^{2}{\tilde{\theta}}\,+\sin^{2}{\tilde{\theta}}\, \right)},  \nonumber\\
g_{IIA'}&=&\frac{m \sin{\tilde{\theta}}\,}{n\cos{\tilde{\theta}}\,\left(\alpha^{2}\cos^{2}{\tilde{\theta}}\,+\sin^{2}{\tilde{\theta}}\, \right) \tilde{R}^{\frac{1}{4}}_{s}A^{\frac{1}{4}}B_{2}}.
\end{eqnarray}
Though the coupling constant $g_{IIB}$ is finite, the coupling constant  $g_{IIA'}$ diverges \footnote{When $n=0$, thus $\theta=0$, we have to replace $(m \sin{\theta}\,)/n$ by $R_{s}/r_{1}$ because $r_{1}$ is then independent of $R_{s}$. This recovers the order estimations of $g_{IIB}\sim \mathcal{O}(R^{1/2}_{s})$ and of $g_{IIA}\sim \mathcal{O}(R^{1/4}_{s})$ of the standard DLCQ procedure discussed in \cite{Seiberg:1997ad}.}. 

\section{$\overline{\text M}$-theory: Multiple M$2$-branes}\label{secviii}

\subsection{a New Planck Length}

For a sensible description of the configuration with the strong coupling $g_{IIA'}$, it is reasonable to go up to the eleven dimensions;
\begin{eqnarray}\label{m2geo}
d\bar{s}^{2}_{M}
&=&\tilde{f}^{-\frac{2}{3}}\left(-d\tilde{t}^{2}+d\bar{x}^{2}_{1} \right)+\frac{\tilde{f}^{\frac{1}{3}}}{\tilde{f} \cos^{2}{\tilde{\theta}}\,+\sin^{2}{\tilde{\theta}}\,}d\bar{x}^{2}_{2}+\tilde{f}^{\frac{1}{3}}\sum^{9}_{i=3}d\tilde{x}^{2}_{i}\nonumber\\
&&+\frac{\tilde{f} \cos^{2}{\tilde{\theta}}\,+\sin^{2}{\tilde{\theta}}\,}{\tilde{f}^{\frac{2}{3}}}\left(d\bar{x}_{11}-\frac{\left(\alpha^{2}-\tilde{f} \right)\cos{\tilde{\theta}}\,\sin{\tilde{\theta}}\,}{\left(\alpha^{2} \cos^{2}{\tilde{\theta}}\,+\sin^{2}{\tilde{\theta}}\, \right)\left( \tilde{f} \cos^{2}{\tilde{\theta}}\,+\sin^{2}{\tilde{\theta}}\,\right)}d\bar{x}_{2} \right)^{2},\nonumber\\
C^{(3)}
&=&\frac{\left(1-\tilde{f} \right)\sin{\tilde{\theta}}\,}{\tilde{f}}d\tilde{t}\wedge d\bar{x}_{11}\wedge d\bar{x}_{1}+\frac{\cos{\tilde{\theta}}\,\left(\tilde{f}-1 \right)}{\tilde{f} \cos^{2}{\tilde{\theta}}\,+\sin^{2}{\tilde{\theta}}\,}d\tilde{t}\wedge d\bar{x}_{1}\wedge d\bar{x}_{2}.
\end{eqnarray} 
Here the coordinate $\bar{x}_{11}$ is compact with the radius
\begin{equation}\label{eq82}
\frac{\bar{r}_{11}}{\bar{l}_{p}}=\frac{\left(\tilde{R}_{11}\bar{r}_{1}\bar{r}_{2} \right)^{\frac{2}{3}}}{\tilde{l}^{2}_{s}},
\end{equation}
where
\begin{equation}\label{newplanck}
\bar{l}_{p}=\left(\tilde{R}_{11}\bar{r}_{1}\bar{r}_{2} \right)^{\frac{1}{3}}.
\end{equation}

This new Planck constant comes as a result of an oblique DLCQ and two successive T-duality transformations on the M-theory configuration. The identification of string T-duality transformations as the transformation of Planck constant in M-theory was first discussed by Susskind \cite{Susskind:1996uh}. We will come back to this point later.

We have a configuration in $\overline{\text M}$-theory characterized by the Planck constant $\bar{l}_{p}$. As $\alpha \rightarrow 0$, the circle size $\bar{r}_{11}$ remains finite as
\begin{equation}
\bar{r}_{11}= \frac{m \sin{\tilde{\theta}}\,}{n\cos{\tilde{\theta}}\,\left(\alpha^{2}\cos^{2}{\tilde{\theta}}\,+\sin^{2}{\tilde{\theta}}\, \right) A B_{2}}\sim \frac{m}{ n\cos{\tilde{\theta}}\,\sin{\tilde{\theta}}\,A B_{2}},
\end{equation}
while the new Planck length shrinks to zero making the $\overline{\text M}$-theory description well-defined;
\begin{equation}
\bar{l}^{3}_{p}=\frac{\sqrt{\tilde{R}_{s}}\,m \sin{\tilde{\theta}}\,}{n\cos{\tilde{\theta}}\,\left(\alpha^{2}\cos^{2}{\tilde{\theta}}\,+\sin^{2}{\tilde{\theta}}\, \right)A^{\frac{5}{2}}B_{2}}\sim\frac{m\sqrt{\tilde{R}_{s}}}{n\cos{\tilde{\theta}}\,\sin{\tilde{\theta}}\, A^{\frac{5}{2}} B_{2}}.
\end{equation}

\subsection{Multiple M$2$-branes}

The geometry describes a number of M$2$-branes spanning the direction $\bar{x}_{1}$ and another direction in  the $(\bar{x}_{2},\,\bar{x}_{11})$-plane while they are smeared over a circle along the residual direction in the same plane. To see this, let us first consider the ADM mass attainable from the above geometry \eqref{m2geo} {\it \'{a} la} Myers-Perry \cite{Myers:1986un}:
\begin{equation}
16\pi \bar{G}_{N}\mu=16\pi \bar{G}^{(8)}_{N}M=5\omega_{6}\tilde{Q}.
\end{equation}
Here, $\omega_{6}=16\pi^{3}/15$ represents the volume of a unit $6$-sphere. We used here the Newton's constant $\bar{G}_{N}$ ($\sim \bar{l}^{9}_{p}$) in $11$-dimensions, thus $\mu$ is the mass density over the directions $\{\bar{x}_{11},\,\bar{x}_{1},\,\bar{x}_{2}\}$. 
On the other hand, the charge is read off from the field strength;  
\begin{equation}
q=\frac{1}{16\pi \bar{G}_{N}}\int *dC^{(3)}=\frac{10\pi \tilde{Q}\left(\bar{r}_{2} \sin{\tilde{\theta}}\,-\bar{r}_{11}\cos{\tilde{\theta}}\, \right)\omega_{6}}{16\pi \bar{G}_{N}}.
\end{equation}
Strictly to say, $q$ is the charge density over two-dimensional spatial world-volume. From the above two equations, we note the relation between the mass density over $3$-volume and the charge density over $2$-volume as 
\begin{equation}
q=2\pi \left(\bar{r}_{2}\sin{\tilde{\theta}}\,-\bar{r}_{11}\cos{\tilde{\theta}}\,\right)\mu.
\end{equation}
This corresponds to BPS relation and suggests that the M$2$-branes are smeared along a circle of radius `$\vert\bar{r}_{2}\sin{\tilde{\theta}}\,-\bar{r}_{11}\cos{\tilde{\theta}} \vert$'. This radius vanishes only when 
\begin{equation}
\alpha^{2}\cos^{2}{\tilde{\theta}}\,+\sin^{2}{\tilde{\theta}}\,=\frac{m}{n}.
\end{equation}
To avoid this singular point in the charge density over $3$-volume, we have assumed at the earlier stage that $m>n$.

\subsection{the Geometric Configuration}

One can rewrite the configuration \eqref{m2geo} in the conventional form of M$2$-branes by introducing new coordinates;
\begin{eqnarray}\label{newcoord}
\left(\begin{array}{c}\bar{X}_{2} \\\bar{X}_{11}\end{array}\right)
&=&\left(\begin{array}{rr}\frac{\alpha^{2}\cos{\tilde{\theta}}\,}{\alpha^{2}\cos^{2}{\tilde{\theta}}\,+\sin^{2}{\tilde{\theta}}\,} & -\sin{\tilde{\theta}}\, \\\frac{\sin{\tilde{\theta}}\,}{\alpha^{2}\cos^{2}{\tilde{\theta}}\,+\sin^{2}{\tilde{\theta}}\,} & \cos{\tilde{\theta}}\,\end{array}\right) \left(\begin{array}{c}\bar{x}_{2} \\\bar{x}_{11}\end{array}\right)\nonumber\\
&=&\left(\begin{array}{rr}\cos{\tilde{\theta}}\, & -\sin{\tilde{\theta}}\, \\\sin{\tilde{\theta}}\, & \cos{\tilde{\theta}}\,\end{array}\right)\left(\begin{array}{cc}1 & 0 \\\frac{(1-\alpha^{2})\cos{\tilde{\theta}}\,\sin{\tilde{\theta}}\,}{\alpha^{2}\cos^{2}{\tilde{\theta}}\,+\sin^{2}{\tilde{\theta}}\,} & 1\end{array}\right)\left(\begin{array}{c}\bar{x}_{2} \\\bar{x}_{11}\end{array}\right).
\end{eqnarray}
In the new coordinates, the geometry becomes
\begin{equation}\label{m2geo}
d\bar{s}^{2}_{M}=\tilde{f}^{-\frac{2}{3}} \left(-d\bar{T}^{2}+d\bar{X}^{2}_{1}+d\bar{X}^{2}_{2} \right)+\tilde{f}^{\frac{1}{3}}\left(\sum^{9}_{i=3}d\bar{X}^{2}_{i}+d\bar{X}^{2}_{11}\right),
\end{equation}
where $\bar{T}=\tilde{t},\,\bar{X}_{1}=\bar{x}_{1},\,\bar{X}_{i}=\tilde{x}_{i}$. Especially as $\alpha$ vanishes, the same transformation gives the standard form of $3$-form field.

\begin{figure}[h]
\begin{center}
\includegraphics[width=6cm]{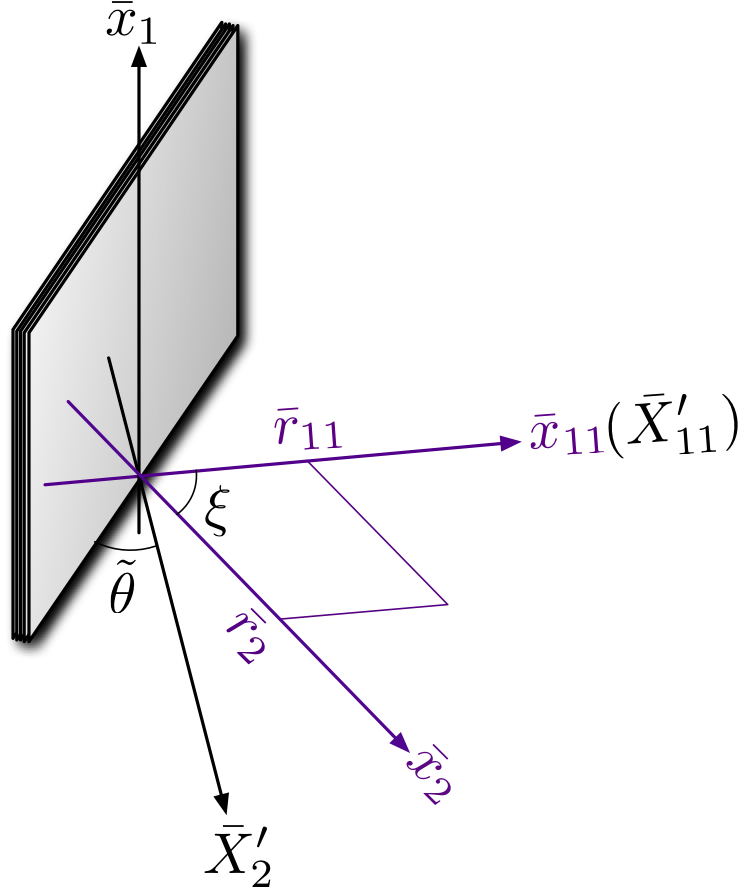}
\caption{A number of multiple M$2$-branes are spanning the direction $\bar{x}_{1}$ and another direction in the $(\bar{x}_{2},\,\bar{x}_{11})$-plane. The coordinates $\{\bar{x}_{2},\,\bar{x}_{11}\}$, forming a slanted torus, are oblique with respect to the orthogonal coordinates $\{\bar{X}'_{2},\,\bar{X}'_{11}\}$.}
\label{m2}
\end{center}
\end{figure}

The Iwasawa decomposition \cite{iwasawa}(the $SL(2,\,\mathbb{R})$ element factorized into an $SO(2,\,\mathbb{R})$ element and a lower triangular matrix with unit diagonal entries) used in the last line of \eqref{newcoord} enables us to figure out the geometrical configuration of these M$2$-branes. The $\bar{X}_{2}$-axis, one of the brane world-volume, is tilted at an angle $\tilde{\theta}$ with respect to the $\bar{X}'_{2}$-axis of an orthogonal frame $(\bar{X}'_{2},\,\bar{X}'_{11})$.  The lower triangular matrix tells us that the frame $(\bar{x}_{2},\,\bar{x}_{11})$ is oblique. The lower off-diagonal element of the triangular matrix can be expressed in terms of the angle $\xi$ that was defined in Eqs. \eqref{angle} and \eqref{angle2};
\begin{equation}
\frac{(1-\alpha^{2})\cos{\tilde{\theta}}\,\sin{\tilde{\theta}}\,}{\alpha^{2}\cos^{2}{\tilde{\theta}}\,+\sin^{2}{\tilde{\theta}}\,}=\cot{\xi}\,.
\end{equation}
We see that the $\bar{x}_{2}$-axis and $\bar{x}_{11}$-axis, which were used in \eqref{m2geo}, are intersecting at the angle $\xi$.
The M$2$-branes are wrapped around this $2$-torus but the direction $\bar{x}_{1}$ unfolds itself in the limit of $\alpha \rightarrow 0$. The size $\bar{r}_{1}$ diverges as $\mathcal{O}(R^{-1/2}_{s})$ while the sizes $\bar{r}_{2}$ and $\bar{r}_{11}$ are of $\mathcal{O}(1)$. Fig. \ref{m2} illustrates the situation. 

\subsection{T-duality in the DLCQ M-theory}

Let us make some remarks on the effect of T-dualities on the DLCQ M-theory. Since the strong coupling $g_{IIA'}$ justifies the uplift to M-theory, it would be meaningful to trace the implication of T-duality transformations in the DLCQ M-theory. We expect some transformation on the eleven dimensional Planck length. Indeed, recasting the right hand side of Eq.~\eqref{newplanck} in terms of the old variables with tilde, one can verify, in the context of supergravity solution, the following relation;  
\begin{eqnarray}\label{tdualm}
\bar{l}^{3}_{p}&=&\tilde{R}_{11}\cdot\frac{\tilde{l}^{2}_{s}\sin{\tilde{\theta}}\,}{\tilde{R}_{s}\left(\alpha^{2}\cos^{2}{\tilde{\theta}}\,+\sin^{2}{\tilde{\theta}}\, \right)}\cdot \frac{\tilde{l}^{2}_{s}}{\tilde{r}_{2}}  \nonumber\\
&=&\frac{\tan^{2}{\tilde{\theta}}\,}{\alpha^{2}\tan^{2}{\theta}\,}\frac{\tilde{l}^{6}_{p}}{\tilde{r}_{1}\tilde{r}_{2}\tilde{R}_{11}}=\frac{\tilde{l}^{6}_{p}}{\tilde{r}_{1}\tilde{r}_{2}\tilde{R}_{11}},
\end{eqnarray}
In the second line, we used the relations
\begin{equation}
\tilde{R}_{s}=r_{1}\sin{\theta}\,=\tilde{r}_{1}\frac{\sin^{2}{\theta}\,}{\sin{\tilde{\theta}}\,},\qquad \frac{1}{\alpha^{2}\cos^{2}{\tilde{\theta}}\,+\sin^{2}{\tilde{\theta}}\,}=\frac{\cos^{2}{\theta}\,}{\alpha^{2}\cos^{2}{\tilde{\theta}}\,}.
\end{equation}
The result \eqref{tdualm} is compatible with the order dependence shown in Tables \ref{table1} and \ref{table2}. We can apparently see that the order difference of $\tilde{r}_{1}$ from $\tilde{r}_{2}$ in the denominator makes the new Planck length $\bar{l}_{p}$ vanishingly small in $\alpha \rightarrow 0$ limit.

T-duality transformation in M-theory has been studied in various contexts. Especially the relation \eqref{tdualm} looks very similar to Eq.~(9) of Ref.~\cite{Susskind:1996uh}, where the corresponding relation was obtained in the context of BFSS M(atrix) theory. (See also Refs.~\cite{Sen:1995cf} \cite{Russo:1996if} \cite{Russo:1997cw} and \cite{Sethi:1997sw}.) 

However, we stress that the above relation \eqref{tdualm} works for the oblique DLCQ M-theory on $\mathbf{T^{2}}$, while the result of Ref.~\cite{Susskind:1996uh} is for BFSS M(atrix) theory on $\mathbf{T^{3}}$. Eq. (9) of Ref. \cite{Susskind:1996uh} can be expressed in our notation as
\begin{equation}\label{tdualsusskind}
\bar{l}^{3}_{p}=\frac{\tilde{l}^{6}_{p}}{\tilde{r}_{1}\tilde{r}_{2}\tilde{r}_{3}}.
\end{equation}
Though both Eqs. \eqref{tdualm} and \eqref{tdualsusskind} involve three directions, the spacelike direction concerning $\tilde{r}_{3}$ of \eqref{tdualsusskind} has been replaced in Eq. \eqref{tdualm} by the M-circle direction concerning $\tilde{R}_{11}$. 

This difference is reminiscent of the extended U-duality group discussed in Ref. \cite{Obers:1998fb}, where it was argued that the U-duality group $E_{p}(\mathbb{Z})$ of DLCQ M-theory on $T^{p}$ should be enhanced to $E_{p+1}(\mathbb{Z})$ if it is unaffected by the lightlike compactification. DLCQ M-theory on $T^{p}$ is nothing but M-theory on $T^{p}\times S^{-}$ where $S^{-}$ is the nearly lightlike circle. Since this latter circle is obtained from the spacelike circle via infinite boosting, it looks reasonable to think that the theory should involve the U-duality group for the M-theory on $T^{p+1}$, that is $E_{p+1}(\mathbb{Z})$, in some way. 

The so-far discussed T-duality transformation in the oblique DLCQ M-theory corresponds to a Weyl generator of the extended U-duality group $E_{3}(\mathbb{Z})=SL(3,\mathbb{Z})\times SL(2,\mathbb{Z})$, but has a slight difference from that of Ref. \cite{Obers:1998fb}. Indeed the relation \eqref{tdualm} is accompanied by the following relations among the compactification sizes;
\begin{equation}\label{udualrel}
\bar{r}_{1}=\frac{\tilde{l}^{2}_{s}}{\tilde{r}_{1}}=\frac{\tilde{l}^{3}_{p}}{\tilde{R}_{11}\tilde{r}_{1}},\qquad \bar{r}_{2}=\frac{\tilde{l}^{2}_{s}}{\tilde{r}_{2}}=\frac{\tilde{l}^{3}_{p}}{\tilde{R}_{11}\tilde{r}_{2}},\qquad \bar{r}_{11}=\frac{\tilde{R}_{11}\bar{r}_{1}\bar{r}_{2}}{\tilde{l}^{2}_{s}}=\frac{\tilde{l}^{3}_{p}}{\tilde{r}_{1}\tilde{r}_{2}}.
\end{equation}
The transformation involves three directions and composes the $\mathbb{Z}_{2}$ sector of the Weyl group $\mathcal{W}(E_{3}(\mathbb{Z}))=\mathbb{Z}_{2}\bowtie \mathcal{S}_{3}$. (Here, the symbol $\bowtie$ implies the group generated by two non-commuting subgroups.) The second factor $\mathcal{S}_{3}$ is nothing but the permutation group of three spacelike directions. In our results, both of the directions concerning $\tilde{R}_{11}=\tilde{r}_{11}$ and $\bar{r}_{11}$ are spacelike while Ref.~\cite{Obers:1998fb}\footnote{In Ref. \cite{Obers:1998fb}, for convenience, the roles of $\tilde{r}_{1}$ and $\tilde{r}_{2}$ were interchanged by the exchange transformation $S_{12}$.} involves the lightlike direction $X'^{-}$ of Eq. \eqref{nulldirection} (in our language).

Consequently, the existence of the above Weyl group strongly suggests that the extendend U-duality can be realized between the $\widetilde{\text{M}}$-theory on an oblique torus (rather than the M-theory on a rectangular torus)  and the $\overline{\text{M}}$-theory on another oblique torus. The U-duality is actually affected by infinitely boosting the spacelike circle of radius $\tilde{R}_{11}$ to the nearly lightlike circle of radius $R$, though it was assumed not in Ref.~\cite{Obers:1998fb}. The lower dimensional Newton constant is not invariant in the DLCQ limit. From the relations \eqref{tdualm} and \eqref{udualrel}, we obtain 
\begin{equation}
\frac{\bar{r}_{1}\bar{r}_{2}\bar{r}_{11}}{\bar{l}^{9}_{p}}=\frac{\tilde{r}_{1}\tilde{r}_{2}\tilde{r}_{11}}{\tilde{l}^{9}_{p}}=\frac{{r}_{1}{r}_{2}{r}_{11}}{\alpha^{7}{l}^{9}_{p}}.
\end{equation}
Seiberg's limit modifies the $8$-dimensional Newton constant of the M-theory (on a rectangular torus) by a rescaling factor, though trivial. Therefore the extended U-duality becomes transparent only after Seiberg's limit has been taken. 

The long chain of duality transformations, taken in this paper, realizes the Weyl generator relating the momentum wave on the internal torus in $\widetilde{\text{M}}$-theory and the multiple M$2$-branes in $\overline{\text{M}}$-theory. Eqs. \eqref{tdualm} and \eqref{udualrel} should give the energy relation for those two states. In order to make sense of the $\overline{\text{M}}$-theory, we devised the oblique DLCQ resulting in the infinite string coupling at the IIA stage. One can see from Eq.~\eqref{metricm} that the $n$ quanta of lightlike momentum wave in $\widetilde{\text{M}}$-theory has the ADM energy
\begin{equation}
P^{0}=P^{11}=\frac{n}{\tilde{R}_{s}}=\frac{n \sin{\tilde{\theta}}\,}{\tilde{r}_{1}}.
\end{equation}
In the last equality, we used Eq.~\eqref{eq52} in $\alpha \rightarrow 0$ limit. Applying the aforementioned U-duality transformation (more precisely its inverse transformation), we obtain
\begin{equation}
\frac{n\bar{r}_{1}\bar{r}_{11}\sin{\tilde{\theta}}\,}{\bar{l}^{3}_{p}}.
\end{equation}
This turns out to be the energy of $n$-tuple of M$2$-branes wrapped over $(\bar{X}_{1},\,\bar{X}_{2})$-directions of Eq.~\eqref{m2geo}, because
\begin{equation}
\bar{X_{1}}=\bar{x}_{1},\qquad \bar{X}_{2}=\bar{x}_{2}\left(\cos{\tilde{\theta}}\,-\cot{\xi}\,\sin{\tilde{\theta}}\, \right)-\bar{x}_{11}\sin{\tilde{\theta}}\,
\end{equation}
and $\bar{X}_{2}\rightarrow -\bar{x}_{11}\sin{\tilde{\theta}}\,$ in DLCQ limit, that is when $\alpha \rightarrow 0$.

\section{Discussions}\label{secix}

We showed that the oblique DLCQ limit on M-theory compactified on a torus $T^{3}$, one of which is the M-circle, is dual to the S-duality transformation of type IIB string theory on $T^{2}$. The deformed torus moduli of M-theory coincide with the transformed vacuum moduli of IIB-theory. The momentum wave propagating along a direction interpolating the M-circle direction and another in $T^{3}$ is dual to multiple $(p,\,q)$-strings of IIB-theory. 

Hence, the coupling of IIA string theory dual to the aforementioned IIB-theory diverges and enhances the non-threshold bound state of D$2$-F$1$, the dual cousin of $(p,\,q)$-strings, to multiple M$2$ branes. 

In Table \ref{table1} and \ref{table2}, we compare the oblique DLCQ and the conventional DLCQ concerning the $\tilde{R}_{s}$-order dependence of various parameters. In the conventional DLCQ on $T^{2}$, there is no notion of $\bar{l}_{p}$ or $\bar{r}_{11}$ because the coupling $g_{IIA'}$ remains finite in $\tilde{R}_{s}\rightarrow 0$ limit.

\begin{table}
\begin{center}
\begin{tabular}[t]{||c|c|c|c|c|c|c||}
\hline\hline \text{parameters} & $g_{IIA}$ & $g_{IIB}$  & $g_{IIA'}$ & $\tilde{l}_{p}$ & $\tilde{l}_{s}$ & $\bar{l}_{p}$ \\\hline \text{Oblique DLCQ} & $\mathcal{O}(\tilde{R}^{\frac{3}{4}}_{s})$ & $\mathcal{O}(1)$ & $\mathcal{O}(\tilde{R}^{-\frac{1}{4}}_{s})$ & $\mathcal{O}(\tilde{R}^{\frac{1}{2}}_{s})$ & $\mathcal{O}(\tilde{R}^{\frac{1}{4}}_{s})$ & $\mathcal{O}(\tilde{R}^{\frac{1}{6}}_{s})$ \\\hline \text{DLCQ} & $\mathcal{O}(\tilde{R}^{\frac{3}{4}}_{s})$ & $\mathcal{O}(\tilde{R}^{\frac{1}{2}}_{s})$ & $\mathcal{O}(\tilde{R}^{\frac{1}{4}}_{s})$ & $\mathcal{O}(\tilde{R}^{\frac{1}{2}}_{s})$ & $\mathcal{O}(\tilde{R}^{\frac{1}{4}}_{s})$ & $\cdot$ \\\hline \hline
\end{tabular}
\end{center}
\caption{For (oblique) DLCQ M-theory on $T^{2}$, the table shows various parameters in the order of $\tilde{R}_{s}$: The second line is for the oblique DLCQ prescription while the third row is for the conventional DLCQ.}
\label{table1}
\end{table}

\begin{table}
\begin{center}
\begin{tabular}[b]{||c|c|c|c|c|c|c||}
\hline\hline \text{radii} & $\tilde{R}_{11}$ & $\tilde{r}_{1}$  & $\tilde{r}_{2}$ & $\bar{r}_{11}$ & $\bar{r}_{1}$ & $\bar{r}_{2}$ \\\hline \text{Oblique DLCQ} & $\mathcal{O}(\tilde{R}_{s})$ & $\mathcal{O}(\tilde{R}_{s})$ & $\mathcal{O}(\tilde{R}^{\frac{1}{2}}_{s})$ & $\mathcal{O}(1)$ & $\mathcal{O}(\tilde{R}^{-\frac{1}{2}}_{s})$ & $\mathcal{O}(1)$ \\\hline\text{DLCQ} & $\mathcal{O}(\tilde{R}_{s})$ & $\mathcal{O}(\tilde{R}^{\frac{1}{2}}_{s})$ & $\mathcal{O}(\tilde{R}^{\frac{1}{2}}_{s})$ & $\cdot$ & $\mathcal{O}(1)$ & $\mathcal{O}(1)$ \\\hline\hline
\end{tabular}
\end{center}
\caption{The sizes of various radii in the order of $\tilde{R}_{s}$. In contrast to the conventional DLCQ, the oblique DLCQ makes the order of $\tilde{r}_{1}$ follow that of $\tilde{R}_{11}$ rather than that of $\tilde{r}_{2}$.}
\label{table2}
\end{table}

\subsection{DLCQ vs. the Oblique DLCQ}

The discrepancy from the conventional DLCQ results comes from the order difference between $\tilde{r}_{1}$ and $\tilde{r}_{2}$, i.e., the rescaled radii of the directions transverse to the DLCQ direction. Table \ref{table1} shows various parameters in the order of $\tilde{R}_{s}$. Though the orders of the fundamental lengths, $\tilde{l}_{p}$ and $\tilde{l}_{s}$, are the same, the order of the coupling constant deviates from that of the conventional DLCQ after the first T-duality toward the IIB-theory. More specifically to say, the string coupling
\begin{equation}
g_{IIA}=\frac{\tilde{R}_{11}}{\tilde{l}_{s}}\sim \mathcal{O}(\tilde{R}^{\frac{3}{4}}_{s})
\end{equation}
acquires a new factor concerning the asymptotic dilaton value at each step of T-duality;
\begin{equation}\label{coupling}
g_{IIB}=\frac{\tilde{R}_{11}}{\tilde{l}_{s}}\frac{\bar{r}_{1}}{\tilde{l}_{s}},\qquad g_{IIA'}=\frac{\tilde{R}_{11}}{\tilde{l}_{s}}\frac{\bar{r}_{1}}{\tilde{l}_{s}}\frac{\bar{r}_{2}}{\tilde{l}_{s}}.
\end{equation}
As we see in Tables \ref{table1} and \ref{table2}, the factor $\bar{r}_{1}/\tilde{l}_{s}$ takes the order
\begin{equation}
\frac{\bar{r}_{1}}{\tilde{l}_{s}}=\frac{\tilde{l}_{s}}{\tilde{r}_{1}}=\begin{cases}
\mathcal{O}(\tilde{R}^{-\frac{3}{4}}_{s})&\qquad\text{Oblique DLCQ}\\
\mathcal{O}(\tilde{R}^{-\frac{1}{4}}_{s})&\qquad\text{DLCQ}
\end{cases}
\end{equation}
while the other factor $\bar{r}_{2}/\tilde{l}_{s}=\tilde{l}_{s}/\tilde{r}_{2}$ is of the same order $\mathcal{O}(\tilde{R}^{-1/4}_{s})$ in both DLCQ's.

\subsection{Parameters of the Oblique DLCQ Prescription}

Let us mention the relation between various parameters we introduced in the oblique DLCQ prescription and those involved in the resulting configuration of the multiple M$2$-branes.

There are five parameters engaged in the oblique DLCQ prescription. The number $n$ representing the momentum sector, the M-circle size $R_{11}$, the sizes $r_1$ and $r_2$ of $T^2$, and the tilting angle $\tilde{\theta}$. In the M$2$ world-volume theory, these parameters are to be encoded into the parameter set composed of the three torus moduli (one Kahler structure modulus $\bar{r}_{11}$ and one set of complex structure modulus $\bar{r}_{2}/\bar{r}_{11},\,\xi$), the size $\bar{r}_1$ of the other compact direction, the charge (number) $\tilde{Q}$ of M$2$-branes. The charge $\tilde{Q}$ of M$2$-branes are related to the mass of those branes via Eq. (8.6). 

Though we do not know at hand how this number is encoded into the world-volume theory of M$2$-branes, it concerns the gauge group $U(n)$ in its corresponding IIA theory, that is, $(2+1)$-dimensional SYM world-volume theory of D$2$-branes. The number $n$ is determined by Eq. (3.22).

\subsection{Decoupling Limit}

In the DLCQ limit, the bulk graviton decouples due to the small Planck length $\bar{l}_{p}\sim \mathcal{O}(\tilde{R}^{1/6}_{s})$. The coupling constant $\bar{\kappa}=\bar{l}^{9/2}_{p}$ becomes weak in the limit. 

Any possible excitation of M$2$-branes is suppressed too in the same limit because a single M$2$-brane has the tension $1/\bar{l}^{3}_{p}$. The DLCQ limit corresponds to the low energy limit. Especially the momentum flow on M$2$-branes is an excitation accompanying the transverse oscillation of the branes. This will break the supersymmetry by half. For example the momentum flow along the $\bar{x}_{1}$-direction in Fig.~\ref{m2} is dual to the momentum flow on multiple $(p,\,q)$-strings in $\widetilde{\text{IIB}}$-theory. This configuration of $(p,\,q)$-helices preserves only $8$ supersymmetries as discussed in Ref. \cite{Cho:2001ys}. 

The interaction with M$5$-branes is also disfavored. First, M$5$-branes, even if they are wrapped on the compact three dimensions of $(\bar{x}_{11},\,\bar{x}_{1},\,\bar{x}_{2})$, are very massive compared to M$2$-branes. The tension $T_{M5}$ of M$5$-branes wrapped on the compact directions will be of the order
\begin{equation}
T_{M5}\sim \frac{\bar{r}_{11}\bar{r}_{1}\bar{r}_{2}}{\bar{l}^{6}_{p}}\sim \mathcal{O}(\tilde{R}^{-1}_{s})T_{M2}.
\end{equation}
 Therefore M$5$-branes cannot be excited energetically by their interaction with M$2$-branes. Second, the supersymmetry is generically broken by their interaction. Even the bound configuration of M$2$- and M$5$-branes can maintain $8$ supersymmetries at most.   

\subsection{Outlook on the World-volume Theory of Multiple M$2$-branes}

Despite the tempered decoupling limit for the M$2$-configuration in $\overline{\text{M}}$-theory, the world-volume theory is involved with the strong Yang-Mills coupling even at the classical level. Since Yang-Mills coupling carries the dimensionality of the inverse length in $(2+1)$-dimensions, it is represented as
\begin{equation}
{g}^{2}_{YM}={g}_{IIA'}\tilde{l}^{-1}_{s}=\frac{\tilde{r}_{11}}{\tilde{r}_{1}\tilde{r}_{2}}=\frac{\bar{r}^{2}_{11}}{\bar{l}^{3}_{p}}.
\end{equation}
Here, we used Eq. \eqref{coupling}. The last equality gives the expression in $\overline{\text{M}}$-theory. With the help of Tables \ref{table1} and \ref{table2}, we note that $g^{2}_{YM}\sim \mathcal{O}(\tilde{R}^{-1/2}_{s})$, thus is divergent in DLCQ limit. In order to avoid this infinite coupling, one could consider taking the other Weyl group elements of the extended U-duality group, that is, the permutations in $\mathcal{S}_{3}$, but it cannot change the situation much. 

We reached a bizarre situation now. At low energy less than $1/\bar{l}_{p}$, the world-volume theory is decoupled from the bulk gravity and is approximated as a field theory. Since the U-duality transformation does not change the supersymmetry, the most promising candidate will be the $\mathcal{N}=8$ super Yang-Mills theory with the adjoint matter fields in $(2+1)$-dimensions. The theory has the infinite Yang-Mills coupling. This latter feature is not new to us because the coupling of $(2+1)$-dimensional Yang-Mills theory, though is well-tempered in the UV regime, runs to the infinity in the IR regime. However in the present theory, we have to consider the infinite Yang-Mills coupling at the classical level.

We do not have a clear view on this theory at hand, but our naive conjecture is that taking the infinite coupling limit on the Yang-Mills theory will remove all the dynamical terms and the theory will become $(2+1)$-dimensional version of the IKKT matrix model type \cite{Ishibashi:1996xs}. Further study should be followed on this and its physical implication.

\acknowledgments
This work is supported by the SRC program of KOSEF through CQUeST with grant number R11-2005-021.

\appendix

\section{Various Periods of the Internal Torus}
In this appendix, we collect the relations among various periods of the internal torus, which are scattered around throughout this paper.
\begin{itemize}
\item{Periods of the rectangular torus in M-theory (Eq. \ref{eq31}):  $(r_{11},\,r_{1},\,r_{2})$} \\
\item{The Effective period of the plane wave propagating on a tilt (Eq. \ref{eq34}):  \\ $R_{s}=r_{11}\cos{\theta}\,$, or $\tilde{R}_{s}=r_{1}\sin{\theta}\,$} \\
\item{Periods of the oblique torus in $\widetilde{\text{M}}$-theory (Eq. \ref{eq315}): $(\tilde{R}_{11},\,\tilde{R}_{1},\,\tilde{r}_{2})$}\\
\begin{eqnarray}
\tilde{R}_{11}&=&\frac{R_{s}}{\cos{\tilde{\theta}}\,}=r_{11}\sqrt{\cos^{2}{\theta}\,+\alpha^{2}\sin^{2}{\theta}\,}\nonumber\\
\tilde{R}_{1}&=&r_{1}\sqrt{\sin^{2}{\theta}\,+\alpha^{2}\cos^{2}{\theta}\,} \nonumber\\
\tilde{r}_{2}&=&\alpha r_{2}\nonumber
\end{eqnarray}
\item{Periods in the asymptotically Minkowskian coordinates in $\widetilde{\text{IIA}}$ theory (Eqs. \ref{eq41}, \ref{eq52}): $(\tilde{r}_{11},\,\tilde{r}_{1},\,\tilde{r}_{2})$}\\
\begin{eqnarray}
\tilde{r}_{11}&=&\tilde{R}_{11}  \nonumber\\
\tilde{r}_{1}&=&r_{1}\sqrt{\sin^{2}{\tilde{\theta}}\,+\alpha^{2}\cos^{2}{\tilde{\theta}}\,}=\frac{\tilde{R}_{s}\left(\alpha^{2}\cos^{2}{\tilde{\theta}}\, +\sin^{2}{\tilde{\theta}}\,\right)}{\sin{\tilde{\theta}}\,}\nonumber\\
\tilde{r}_{2}&=&\alpha r_{2}\nonumber
\end{eqnarray}
\item{Periods in $\widetilde{\text{IIA}'}$ theory (Eqs. \ref{dualR1},\,\ref{dualR2}): $(\tilde{r}_{11},\,\bar{r}_{1},\,\bar{r}_{2})$ 
\begin{eqnarray}
\bar{r}_{1}&=&\frac{\tilde{l}^{2}_{s}}{\tilde{r}_{1}}  \nonumber\\
\bar{r}_{2}&=&\frac{\tilde{l}^{2}_{s}}{\tilde{r}_{2}}  \nonumber\\
\end{eqnarray}}
\item{Periods in $\overline{M}$-theory (Eqs. \ref{eq82}, \ref{newplanck}): $(\bar{r}_{11},\,\bar{r}_{1},\,\bar{r}_{2})$}
\begin{eqnarray}
\frac{\bar{r}_{11}}{\bar{l}_{p}}&=&\frac{\left(\tilde{r}_{11}\bar{r}_{1}\bar{r}_{2}\right)^{\frac{2}{3}}}{\tilde{l}^{2}_{s}}  \nonumber\\
\bar{l}_{p}&=&\left(\tilde{r}_{11}\bar{r}_{1}\bar{r}_{2}\right)^{\frac{1}{3}}  \nonumber\\
\end{eqnarray}
\end{itemize}

\end{document}